\documentstyle[12pt,titlepage]{article}
\font\grande=cmr10 scaled \magstep4
\font\medio=cmr10 scaled \magstep2

\def\laq{\raise 0.4ex\hbox{$<$}\kern -0.8em\lower 0.62
ex\hbox{$\sim$}}
\def\gaq{\raise 0.4ex\hbox{$>$}\kern -0.7em\lower 0.62
ex\hbox{$\sim$}}

\begin{document}
\bibliographystyle {unsrt}
\newcommand{\pa}{\partial}

\titlepage
\begin{flushright}
CERN-TH/95-199 \\
DFTT 44/95
\end{flushright}
\vspace{15mm}

\begin{center}
{\grande Constraints on\\

\     \\

Supergravity Chaotic Inflationary Models}\\
\vspace{5mm}
{\grande }
\vspace{10mm}
M. C. Bento\footnote{On leave fom Departamento de F\'{\i}sica,
  Instituto Superior T\'ecnico, Av. Rovisco Pais, 1096 Lisboa Codex.} \\
{\em Theory Division, CERN, CH-1211 Geneva 23, Switzerland} \\
and\\
O. Bertolami$^{1,}$\footnote{Also at Theory Division, CERN.} \\
{\em INFN-Sezione Torino, Via P. Giuria 1, 10125 Turin,
Italy} \\
\end{center}

\vspace{10mm}
\centerline{\medio  Abstract}
\noindent
We discuss, in the context of $N=1$ hidden sector non-minimal
supergravity chaotic inflationary models,
constraints on the parameters of a polynomial superpotential
resulting from existing bounds on the reheating temperature
and on the amplitude of
the primordial energy density fluctuations as inferred from COBE.
We present a specific two-parameter chaotic inflationary model
which satisfies these constraints and discuss a possible scenario
for adequate baryon asymmetry generation.
\vspace{5mm}
\vfill
\begin{flushleft}
CERN-TH/95-199 \\
DFTT 44/95 \\
September 1995 \end{flushleft}

\newpage

\setcounter{equation}{0}

Recently, a great deal of attention has been devoted to  possible
implementations of
the inflationary scenario in supergravity/superstring models.  Besides
ensuring that there is sufficient inflation
to solve the initial condition problems of the standard cosmological model,
these models have to explain observational limits such as the magnitude
of the energy density perturbations required to explain the
anisotropies in
the Cosmic Microwave Background radiation observed by COBE
\cite{cobe}. Furthermore, in supergravity cosmological models, the reheating
temperature should not
exceed $T_{RH}~\laq~ 2.5\times  10^8 (100~\hbox{GeV}/m_{3/2})$ GeV
\cite{ellis},  not to generate an abundance of
gravitinos which would photo-dissociate light elements produced in primordial
nucleosynthesis. In the context of superstring cosmology,
inflationary models have to face further problems such as the fate of
the dilaton and moduli fields and the so-called post-modern Polonyi problem
\cite{coughlan}. Although many of these issues can be addressed in a
simple chaotic model where the dilaton plays the role of the inflaton
and its potential is dominated by  quadratic and/or  quartic self-couplings
\cite{bento}, this is not the case for the dilaton
potentials generated by the  supersymmetry-breaking
mechanisms currently preferred in superstring-based models,
i.e. gaugino condensation; moreover, these  potentials appear not to
be sufficiently flat to allow inflation to occur \cite{stein}.

In this letter, we study constraints on  $N = 1$  non-minimal supergravity
realizations of chaotic inflationary
models inspired in the superstring, resulting from the abovementioned
cosmological bounds. Chaotic inflationary models \cite{linde}
stand out as the most natural ones in what the initial
conditions for the onset of inflation are concerned, particularly in the
context of
supergravity and superstring theories, where the natural scale for fields is
the Planck scale. Realizations
of chaotic inflation in minimal and in $SU(1,1)\ N=1 $ supergravity
theories have been studied in \cite{gon},\cite{gon1}.  We
discuss a specific model, with a two-scale
chaotic inflationary sector, which could originate from
the existence of two gaugino condensation and/or gauge symmetry
breaking scales, that  can accommodate in a satisfactory
way the bounds on the reheating temperature and energy density
fluctuations. It follows from our analysis that,  as first pointed out
in Ref. \cite{ross}, a chaotic
inflationary model requires more than one scale to reproduce the abovementioned
constraints; this is essentially due to the fact that in chaotic
models the
slow roll-over period  occurs  around the Planck
scale and not, as in
Refs. \cite{ross},\cite{hrross}, some orders of magnitude below.

We shall assume that the inflaton is the scalar component of a gauge
singlet superfield, $\Phi$, in the hidden sector of the theory. We start
by  splitting
the superpotential in a supersymmetry-breaking, a
gauge and an inflationary part, as suggested in \cite{ross},\cite{hrross}:

\begin{equation}
\label{a}
W= P + G + I.
\end{equation}

The scalar potential for the inflaton field is
obtained from the superpotential $I(\Phi)$ as \cite{crem}

\begin{equation}
\label{aa}
V(\phi)=\exp (g)\left. |\frac{\partial g}{\partial \Phi}|^{2} (
\frac{\partial^{2} K}{\partial \Phi \partial \Phi^{*}})^{-1}
\right|_{\Phi=\phi}.
\end{equation}
where a suitable choice for the Kaehler function in $SU(1,1)$ supergravity is
the following \cite{gon1}:

\begin{equation}
\label{aa1}
K(\Phi, \Phi^{*})= - \frac{3}{2}~\ln |h(\Phi, \Phi^{*})|^2 + g(\Phi, \Phi^{*})
\end{equation}
with

\begin{equation}
\label{aa1}
g(\Phi, \Phi^{*})= (\frac{\Phi - \Phi^{*}}{M})^2 + \ln |\frac{I(\Phi)}{M^3}|^2
\end{equation}
and $\frac{\partial^{2} h(\Phi, \Phi^{*})}{\partial
\Phi \partial \Phi^{*}} = \mu$, $\mu$ being a constant and
$M=M_{P}/\sqrt{8 \pi}$, where $M_P$ is the Planck mass.
Requiring the cosmological constant to vanish and that supersymmetry
remains unbroken at the minimum of the
potential,  $\Phi=\Phi_o$,  leads to
the following constraints on the superpotential:

\begin{equation}
\label{ac}
I(\Phi_o)=\frac{\partial I}{\partial \Phi}(\Phi_o)=0.
\end{equation}

Consider the most general polynomial  superpotential

\begin{equation}
\label{aba}
I(\Phi)=\sum_{n=0} \frac{a_n}{M^{n-2}}\Phi^n,
\end{equation}
where the $a_n$ are mass parameters. Dropping the linear
and  the non-renormalizable terms $(n>3)$ in (\ref{aba}), the latter  leading
to
too large tensor perturbations of the microwave background \cite{enqvist},
we are left with

\begin{equation}
\label{ab}
I(\Phi)= I_o +  a~\Phi^2 + \frac{b}{M}~\Phi^3~,
\end{equation}
where $I_o$ is a constant and $a$ is positive.
The conditions (\ref{ac}), applied to the superpotential of
Eq.~(\ref{ab}), give two solutions for $ \phi_o, I_o$:

\begin{eqnarray}
(\phi_o,\ I_o)  &= &( 0,~ 0 ); \\
                    &= & (-2 a M/3 b,\ -4 a^3 M^2/27 b^2).
\end{eqnarray}

We shall first consider the case $\phi_o=I_o=0$. The relevant part of the
inflaton
potential (along the real $\phi$ direction) is then given by:
\begin{eqnarray}
\label{aca}
V(\phi)& = & \mu^3 M^2 \left[ 4~a^2 \left(\frac{\phi}{M}\right)^2
            + 12~a~b \left(\frac{\phi}{M}\right)^3
            + 9~b^2 \left(\frac{\phi}{M}\right)^4\right].
\end{eqnarray}

In the chaotic inflationary scenario, the  scalar field starts rolling towards
its minimum at the origin from an initially large value beyond the Planck
scale. During this process, the domains of the Universe filled with a
sufficiently homogeneous $\phi$ field expand according to the Friedmann
equation

\begin{equation}
\label{acb}
H^2=\frac{1}{3 M^2} \left(\frac{1}{2}{ \dot\phi}^2 + V\right),
\end{equation}
and the $\phi$ field evolves according to

\begin{equation}
\label{acc}
\ddot \phi + 3 H \dot \phi  + \frac{dV}{d\phi}=0.
\end{equation}

In the region $|\phi|>~\hbox{few} ~M_P$, the field rolls down very slowly and
the
terms $\dot\phi^2$ in (\ref{acb}) and $\ddot \phi$ in (\ref{acc}) can be
neglected. In this
region, the potential $V(\phi)$ can be approximated by

\begin{equation}
\label{acd}
V(\phi)\approx \frac{9b^2}{M^2} \phi^4.
\end{equation}

The total number of $e$-folds of inflation is given by

\begin{equation}
\label{ad}
N\equiv \ln \frac{a(\phi_{e})}{a(\phi_{i})}= -
\frac{1}{M^2}\int_{\phi_{i}}^{\phi_{e}} \frac{V}{V'} d\phi \approx
\frac{\pi}{M_P^2} \left( \phi_i^2 - \phi_e^2\right)  .
\end{equation}

Hence, it is required that $\phi_i\gaq 8.3M_P$, for $\phi_e \approx M_P$,
to get $N\gaq~ 65$.

After inflation,  the field $\phi$ begins to
oscillate about its minimum, thus reheating the Universe. At minimum,
the inflaton field has a mass (from here onwards we set $\mu = 1$)

\begin{equation}
\label{ba}
m_\phi= 2 \sqrt{2} ~a.
\end{equation}

Since the inflaton is hidden from the other sectors of the theory, it
couples to lighter fields with strength $\sim a/M$, leading to a decay width

\begin{equation}
\label{bb}
\Gamma_\phi \sim \frac{m_\phi}{(2\pi)^3} \left(\frac{a}{M}\right)^3,
\end{equation}
and a reheating temperature

\begin{equation}
\label{bc}
T_{RH}\approx \left(\frac{30}{\pi^2 g_{RH}}\right)^{1/4} \sqrt{M \Gamma} \sim
\frac{2}{\pi^2}\left(\sqrt{\frac{15}{g_{RH}}}\frac{a^3}{M}\right)^{1/2},
\end{equation}
where $g_{RH}$ is the number of degrees of freedom at $T_{RH}$.
Notice that, as $\Gamma_\phi << m_\phi$ and bilinear couplings are
absent\footnote{We thank Andrei Linde for discussions on this subject.},
parametric resonance effects \cite{kofman} are not important in this case
(see also Ref. \cite{bross} for a discussion on the effect of the presence
of anharmonic couplings between chiral superfields in the reheating process).

As mentioned above, a severe upper bound on $T_{RH}$ comes from the
requirement that sufficiently few gravitinos are regenerated in the
post-inflationary reheating epoch. Indeed, once regenerated beyond a
certain density, stable thermal
gravitinos would dominate the energy density of the Universe or, if
they decay, have undesirable effects on nucleosynthesis and  light element
photo-dissociation and lead to distortions
in the microwave background.
This implies  the following bounds\footnote
{Fischler \cite{fisch} suggested that heat-bath effects might
greatly enhance  the gravitino regeneration rate at high temperature
and thereby lower the bound on $T_{RH}$, a
claim that has since been questioned \cite{leigh}.} \cite{sarkar}:

\begin{equation}
\label{bca}
T_{RH}~\laq ~2\times 10^9,~6\times
10^9~\hbox{GeV}~~~~\hbox{for}~~~~m_{3/2} = 1,~10~\hbox{TeV}.
\end{equation}

For our model, demanding that $T_{RH}$ be less than $2\times 10^9$ GeV then
leads, for $g_{RH} \approx 150$, to a bound on parameter $a$ as

\begin{equation}
\label{bd}
\frac{a}{M}~\laq~3.7\times 10^{-6}.
\end{equation}

We have checked that the above bound also ensures that gravitino production via
inflaton decay is sufficiently
suppressed, as in \cite{ross}.

Further constraints on the parameters of the superpotential
can be derived  from the spectrum of adiabatic density
fluctuations, which is given, in terms of the potential, by \cite{lyth}:

\begin{equation}
\label{ca}
\delta_H\equiv \sqrt{4 \pi} \left( \frac{\delta \rho}{\rho}\right)_H =
  \frac{1}{5\sqrt{3}\pi M^3}\frac{V_{\star}^{3/2}}{V_{\star}^\prime},
\end{equation}
where the subscript $\star$ indicates that the right-hand side should
be evaluated as the comoving scale $k$ equals the Hubble radius $(k=a
H)$ during inflation. Neglecting  higher multipoles in  the Cosmic
Microwave Background radiation observed by COBE, the best fitting quadrupole
moment obtained
from the angular power spectrum  corresponds to \cite{cobe}

\begin{equation}
\label{cb}
\delta_H\approx 2.3\times 10^{-5}.
\end{equation}
with an uncertainty of $\sim 10\%$.
Combining (\ref{ca}) and (\ref{cb}), we obtain for our model

\begin{equation}
\label{cc}
\delta_H \approx \frac{4.25\times 10^{2}}{5\sqrt{3}~\pi~M}\frac{(4~a^2 +
2.47\times 10^{2}~a~b
+ 3.83\times 10^{3}~b^2)^{3/2}}{(8~a^2 + 7.42\times 10^{2}~a~b
+ 1.53\times 10^{4}~b^2)},
\end{equation}
which, using the constraint on $a$ derived above, Eq.~(\ref{bd}), implies,
in turn, a bound on parameter $b$

\begin{equation}
\label{cd}
\frac{b}{M}~\laq~- 2.5\times 10^{-7}.
\end{equation}

%Note that, for  an accurate analysis, one would have to include the
%effect of spectral tilt and gravitational waves on the COBE
%normalization \cite{}. However, such changes are not significant in
%the present context and for simplicity we shall not include them here.

For the other solution of Eq.~(\ref{ac}), i.e. $ \phi_o =- 2 a M /3 b,\
I_o = - 4 a^3 /27 b^2 $, we obtain  the same bounds on $a$ and $b$, as
expected.

 Of course, our results would be modified if, instead of (\ref{bca}),
 there were
 stricter bounds on the
 reheating temperature \cite{fisch}. For instance, for $T_{RH}~\laq~10^6$ GeV,
we  obtain:

\begin{equation}
\label{ce}
\frac{a}{M}\laq ~2.3\times 10^{-8}~,~ \frac{b}{M}\laq~9.5\times 10^{-8} .
\end{equation}

A realistic scenario for baryogenesis can be built considering the decay of
the inflaton into the matter field states in the gauge sector of the
superpotential (1). As the coupling between the inflaton and these states is
only gravitational, the former will decay into the heaviest states available
\cite{hrross}, which
will then generate the baryon asymmetry through decays into quarks and leptons.
The baryon-antibaryon number density will then be given essentially in terms
of the asymmetry following from inflaton decay:

\begin{equation}
\label{cf}
n_{B- \bar B}~\approx~n_{\phi}~ \delta B~,
\end{equation}
where $\delta B$ is the baryon asymmetry generated per decay.
The photon number density can be given as a function of
the inflaton density and the reheating temperature

\begin{equation}
\label{cg}
n_{\gamma}~\approx~\frac{\rho_{\phi}}{T_{RH}}~\approx~\frac{n_{\phi}m_{\phi}}
{T_{RH}}~,
\end{equation}

\noindent
so that, from (13), (16) and (17), the asymmetry can be expressed as

\begin{equation}
\label{ch}
\xi \equiv
\frac{n_{B- \bar
    B}}{n_{\gamma}}~\approx~\frac{T_{RH}}{m_{\phi}}~\delta B \sim
10^{-4}~\delta B~,
\end{equation}

\noindent
which allows us to obtain the observed value, $\xi \sim 10^{-10}$,
provided $\delta B$ has a suitable value. Although we shall not try to
specify here  how, in a
concrete particle physics  model, the required value for $\delta B$
could be produced radiatively,  we stress
that the asymmetry can be created even though the reheating temperature is
as low as or lower than the bound (16).
 We also point out that, in models such as the ones discussed in
Ref.~\cite{bento} (see also \cite{linde1}), the inflaton  (the dilaton,
in that instance) is directly
coupled to a GUT Higgs field
and the mechanism discussed above can be easily implemented. This GUT Higgs
field can be endowed with a suitable potential that may allow a subsequent
period of inflation.

Of course, other scenarios  could be envisaged to
generate the baryon asymmetry, which can be completely or fairly
independent of
inflaton decay, depending on whether or not this decay dilutes
the generated baryon
asymmetry (see e.g. the  second reference in [4]). An example is
the Affleck-Dine mechanism
\cite{affleck}, recently implemented in the
context of supergravity string-inspired models \cite{drt}.
The main feature invoked in
these models is that, during inflation, supersymmetry-breaking soft terms, with
mass terms of the order of the Hubble parameter, are naturally induced
\cite{dfn}. Since the cosmology of string theories has problems
associated with the fate of the moduli fields, an additional
period of late inflation seems to be a rather natural way to avoid the problems
associated with the presence of these fields \cite{rt}. The possibility that
this late period of inflation is related with baryon asymmetry generation
itself is certainly very appealing. In the context of our model, an
Affleck-Dine baryogenesis scenario like the one in \cite{drt} can also
be constructed. Non-renormalizable terms, together with
soft supersymmetry-breaking terms arising from  a second period of GUT
inflation, give rise, along some direction in the space of scalar fields
that carry
baryon and lepton number ($\chi$), such as squarks and sleptons,  to
the potential \cite{drt},\cite{dfn}:

\begin{equation}
\label{ci}
V(\chi) \approx c~ H^2 |\chi|^2 + a~ \lambda \frac{H |\chi|^n}{n M^{n-3}} +
\lambda^2 \frac{|\chi|^{2n-2}}{M^{2n-6}}~~,
\end{equation}
 where $a$ and $c$ are O(1) constants -- ``$a$ terms'' are important
for $B$ and $L$
violation -- and $\lambda$ is a coupling constant.
This potential admits a non-trivial minimum
$|\chi_0| \approx \left({\frac{-c}{n-1}}\right)^{1/2}\frac{H}{\lambda}~M^{n-3}$
, for $c < 0$.
After the second period of inflation, when $H \approx m_{3/2}$, the field
oscillates around $\chi_0$
and a baryon asymmetry such as (\ref{ch}) is generated, with $\delta B$
given essentially by $(\chi_0/M_P)^2$ \cite{drt}.

Finally, we comment on possible origins for the (two) scales of our
model. Let us first consider the possibility that these scales are
induced by gauge symmetry breaking. In fact, as suggested in Ref.
\cite{ross}, once gauge non-singlet fields, $\Psi$, acquire a v.e.v. along a
$D$-flat direction, thereby breaking the gauge symmetry, a v.e.v.
for the massive gauge singlet fields coupled to them will be induced;
these v.e.v.'s then feed through to the inflationary sector via
couplings between the latter fields and the inflaton, leading
to a superpotential of the form \cite{ross}

\begin{equation}
\label{da}
I(\Phi)=a~M^2~f\left( \frac{\Phi}{M}\right),
\end{equation}
where $a=\langle\Psi\rangle^2\langle\bar \Psi\rangle^2/ M^3$ and $f(x)$
is a polynomial function.
For $\langle\Psi\rangle\sim 10^{16}$ GeV,
we obtain $a\sim 10^{10}$ GeV. Of course, for our model, at least two
operators coupling $\Psi$ and $\Phi$ fields would be required, whose form is
determined by e.g. discrete symmetries, which arise naturally   in the
context of string-inspired phenomenological models as a
consequence of the symmetries of the compactification manifold \cite{ross3}. In
fact, one
hopes that  these and/or other symmetries present in the fundamental
theory may also explain the absence of higher-order non-renormalisable terms,
which are not
small in chaotic inflationary models, where $\Phi\geq M_P$.
Other  possibilities for the origin of these scales is that there are
two stages of gauge symmetry breaking or that they arise from
the supersymmetry-breaking sector, through the gaugino condensation of two
distinct subgroups of the hidden group $E_8$.

Let us now summarize our results. We have shown that, in order to satisfy COBE
data and to keep the reheating temperature sufficiently low not to regenerate
an excessive abundance of gravitinos, non-minimal supergravity
chaotic inflationary models require at
least two independent scales in the superpotential. These scales, which can
be related with the scales of gaugino condensation and/or gauge symmetry
breaking, are significantly below the Planck scale. We have also analysed
how inflaton decay can
potentially explain the observed baryon asymmetry of the Universe.

\vspace{1cm}
\noindent
{\bf Acknowledgement}

\vspace{0.2cm}
\noindent
We would like to thank Graham Ross for important discussions and Subir Sarkar
for various suggestions.

\end{document}